\documentclass[aps,prb,reprint,showpacs,superscriptaddress]{revtex4-1}
\usepackage{graphicx}
\usepackage{latexsym}
\usepackage[T1]{fontenc}
\usepackage{amsmath}
\usepackage{endnotes}
\input{epsf}

\begin{document}

\preprint{APS/123-QED}

\date{\today}

\title{Thermodynamic phase diagram and phase competition in BaFe$_2$(As$_{1-x}$P$_x$)$_2$ studied by thermal expansion}

\author{A. E. Böhmer}
\affiliation{Institut für Festkörperphysik, Karlsruhe Institute for Technology, 76021 Karlsruhe, Germany}
\affiliation{Fakultät für Physik, Karlsruhe Institute for Technology,  76131 Karlsruhe, Germany}

\author{P. Burger}
\affiliation{Institut für Festkörperphysik, Karlsruhe Institute for Technology, 76021 Karlsruhe, Germany}
\affiliation{Fakultät für Physik, Karlsruhe Institute for Technology,  76131 Karlsruhe, Germany}

\author{F. Hardy} 
\affiliation{Institut für Festkörperphysik, Karlsruhe Institute for Technology, 76021 Karlsruhe, Germany}

\author{T. Wolf}
\affiliation{Institut für Festkörperphysik, Karlsruhe Institute for Technology, 76021 Karlsruhe, Germany}

\author{P. Schweiss}
\affiliation{Institut für Festkörperphysik, Karlsruhe Institute for Technology, 76021 Karlsruhe, Germany}

\author{R. Fromknecht}
\affiliation{Institut für Festkörperphysik, Karlsruhe Institute for Technology, 76021 Karlsruhe, Germany}

\author{H. v. Löhneysen}
\affiliation{Institut für Festkörperphysik, Karlsruhe Institute for Technology, 76021 Karlsruhe, Germany}
\affiliation{Physikalisches Institut, Karlsruhe Institute for Technology, 76131 Karlsruhe, Germany}

\author{C. Meingast}
\affiliation{Institut für Festkörperphysik, Karlsruhe Institute for Technology, 76021 Karlsruhe, Germany}

\author{H. K. Mak}
\affiliation{The Hong Kong University of Science \& Technology, Clear Water Bay, Kowloon, Hong Kong}

\author{R. Lortz}
\affiliation{The Hong Kong University of Science \& Technology, Clear Water Bay, Kowloon, Hong Kong}

\author{S. Kasahara}
\affiliation{Research Center for Low Temperature and Materials Sciences, Kyoto University, Kyoto 606-8501, Japan}

\author{T. Terashima}
\affiliation{Research Center for Low Temperature and Materials Sciences, Kyoto University, Kyoto 606-8501, Japan}

\author{T. Shibauchi}
\affiliation{Department of Physics, Kyoto University, Kyoto 606-8502, Japan}

\author{Y. Matsuda}
\affiliation{Department of Physics, Kyoto University, Kyoto 606-8502, Japan}

\begin{abstract}
High-resolution thermal-expansion and specific-heat measurements were performed on single 
crystalline BaFe$_2$(As$_{1-x}$P$_x$)$_2$ ($0\leq x \leq0.33$, $x=1$). The observation of clear anomalies 
allows to establish the thermodynamic phase diagram which features a small coexistence region of SDW and 
superconductivity with a steep rise of $T_c$ on the underdoped side. Samples that undergo the tetragonal-orthorhombic 
structural transition are detwinned \textit{in situ}, and the response 
of the sample length to the magneto-structural and superconducting transitions 
is studied for all three crystallographic directions. It is shown that a reduction of the magnetic order by superconductivity 
is reflected in all lattice parameters. On the overdoped side, superconductivity affects the lattice parameters 
in much the same way as the SDW on the underdoped side, suggesting an intimate relation between the two types of order. 
Moreover, the uniaxial pressure derivatives of $T_c$ are calculated using the Ehrenfest relation and 
are found to be large and anisotropic. A correspondence 
between substitution and uniaxial pressure is established, i.e., uniaxial pressure along the $b$-axis ($c$-axis) corresponds to a decrease (increase) of the P content.
By studying the electronic contribution to the thermal expansion we find 
evidence for a maximum of the electronic density of states at optimal doping. 
\end{abstract}

\pacs{74.70.Xa, 74.25.Bt, 74.62.Dh, 74.62.Fj}

\maketitle

\section{Introduction}\label{sec:intro}

In the intensively studied 122 family of iron-based superconductors, 
superconductivity can be induced by various substitutions in BaFe$_2$As$_2$ (Ba122), e.g., K for Ba, Co or Ru for Fe and also P for 
As, as well as by hydrostatic pressure\cite{Rotter2008,Sefat2008,Sharma2010,Jiang2009,Kasahara2010,Alireza2009}. 
In all cases, the resulting phase diagrams are surprisingly 
similar: a superconducting dome arises when the magneto-structural 
transition of the parent compound is suppressed by either increasing doping levels or 
increasing pressure. The isovalent substitution of As by P in BaFe$_2$(As$_{1-x}$P$_x$)$_2$ (P-Ba122) with a 
maximum $T_c$ of $30\,$K is particularly interesting since 
(nominally) no additional charge carriers are introduced. Observation of the de 
Haas-van Alphen (dHvA) effect for $0.41<x<1$ (Refs. \onlinecite{Shishido2010,Analytis2010}) demonstrates the low 
scattering rate of the defects introduced by P substitution. Similarly, the absence of 
quasiparticle scattering by the dopant atoms in this system was demonstrated in a study on vortex pinning\cite{vanderBeek2010}. 
It was also found that the phase diagram of P-Ba122 can be tracked with hydrostatic pressure from any starting 
P concentration\cite{Klintberg2010}, suggesting that pressure is somehow equivalent to doping, 
as has also been found for the Co-doped system\cite{Drotziger2011,Meingast2011}.  Evidence for a quantum critical point close to 
optimal doping was obtained from non-Fermi liquid transport\cite{Kasahara2010}, nearly zero Curie-Weiss temperature in 
NMR\cite{Nakai2010II} and mass enhancement in dHvA measurements\cite{Shishido2010}.

\begin{figure*}[t]
\begin{center}
\includegraphics[width=160mm]{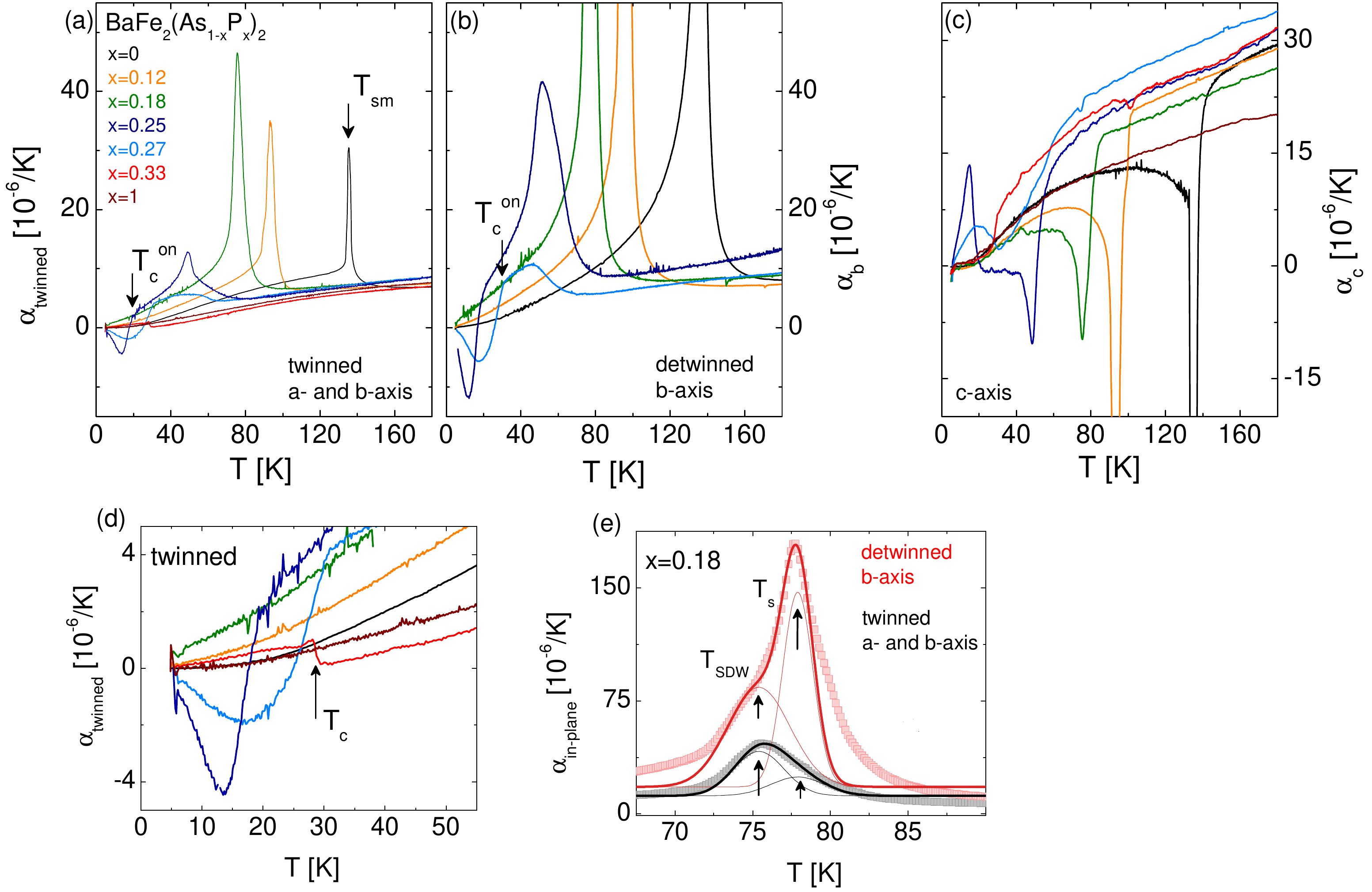}
\end{center}
\caption{Uniaxial thermal-expansion coefficients $\alpha_i$ as a function of temperature $T$ for (a) 
twinned in-plane and (b) detwinned measurements (along the $b$-axis) and 
(c) along the $c$-axis for various substitution levels as indicated in panel 
(a). Arrows indicate examples of the temperature of the magneto-structural 
transition $T_{sm}$ and $T_c^{\ on}$. (d) Magnified view of the low-temperature 
region of (a). (e) shows an example of how the splitting of the magneto-structural 
transition into a structural transition at $T_s$ and a magnetic transition at $T_{SDW}$ is inferred from the data.}
\label{fig:alphaall}
\end{figure*}

Here, we present a study of thermodynamic properties of the P-Ba122
 system focusing on the underdoped and slightly 
overdoped regime ($0\leq x \leq0.33$) using thermal-expansion, specific-heat and magnetization measurements. Thermal 
expansion of single crystals probes the uniaxial pressure dependence of the entropy and is 
therefore well suited to study the effect of pressure on superconductivity and spin-density-wave (SDW) ordering.
Uniaxial pressure effects are expected to be quite important in this class of 
materials due to their anisotropic crystal structure\cite{Hardy2009,Budko2009,Meingast2011}. 
High-resolution thermal-expansion measurements also provide an ideal method for determining 
the thermodynamic phase diagram of these systems. We find that the phase 
diagram of P-Ba122 resembles those of other 122 superconductors, with the 
exception of a much steeper rise of the bulk $T_c$ on the underdoped side. We show that the small pressure applied 
by the dilatometer on the sample is sufficient to detwin the crystals \textit{in situ}. 
By combining the results of crystals in the twinned and detwinned states, we show that the 
orthorhombicity of underdoped samples is reduced by the onset of 
superconductivity, which is a clear sign of the competition between superconducting and SDW/structural order parameters.
Moreover, we establish a linear relation between P substitution and uniaxial pressure, 
as far as their effects on the phase diagram are concerned. 
Finally, an evaluation of the electronic component of the thermal expansion 
reveals an increasingly pressure sensitive density of states on approaching 
optimal doping from the underdoped side and a maximum of the Sommerfeld coefficient at optimal doping.
Throughout this text, the term ``doping'' will be used in a broad sense, as an equivalent to substitution. 

The paper is organized as follows. In section II and III we present the experimental 
details and results, from which we construct the thermodynamic phase diagram (section IV).  
In section V we discuss the interplay of the structural/magnetic order with superconductivity.  
The uniaxial pressure effects on $T_c$ and the electronic density of states 
are presented in Section VI, and conclusions are provided in Section VII. A brief preliminary 
report\cite{Boehmer2011} on parts of this work has been presented previously.

\section{Experimental details}\label{sec:experimental}

Single crystals of P-substituted Ba122 were grown from stoichiometric mixtures 
of the starting materials as described in Ref.~\onlinecite{Kasahara2010}. 
Crystals of pure BaFe$_2$P$_2$ were grown from self-flux using an Al$_2$O$_3$ crucible in a closed steel container. 
Ba and pre-reacted FeP were mixed in a ratio of 1:5, heated up to 1300$^\circ$C and slowly cooled down to 1200$^\circ$C at a rate of 0.3$^\circ$C/h. 
The typical crystal size is $\sim500\times500\times100\,\mu$m$^3$. 4-circle x-ray 
structural refinements were conducted on three of our samples and yielded a P content
 of $x=0.25(1), 0.30(1), 0.33(1)$. The P content of the other samples 
($x=0.12(2), 0.18(2), 0.26(2), 0.27(2)$) was determined by energy-dispersive 
x-ray analysis (EDX) on these samples and complemented by 4-circle x-ray diffraction on samples from the 
same batch, hence the larger error.  

Thermal expansion was measured in a home-built capacitive dilatometer with a 
typical resolution of $0.1-0.01$\,\AA\ (Ref. \onlinecite{Meingast1990}). In the dilatometer, the sample is pressed 
against one plate of a plate-type capacitor (with a force of $F\approx 0.2\,$N) 
so that a change of the sample length results in a changing capacitor gap. 
Samples were mechanically detwinned \textit{in situ} by mounting them such that the 
dilatometer pressure is directed along their tetragonal [110] 
direction\cite{Fisher2011}. In this configuration, thermal expansion along the (shorter) 
orthorhombic $b$-axis is measured. Comparison with twinned samples, for which the 
dilatometer pressure is applied along the tetragonal [100] direction, allows to 
estimate the thermal expansion along the orthorhombic $a$-axis as well. Accurate
 data could be obtained in spite of the (for dilatometry) extremely 
small sample size. 

The specific heat on the tiny samples with masses in the order of 100 micrograms 
was measured with a home-made micro-relaxation calorimeter using a `long 
relaxation' technique \cite{vanHeumen2007}. Each relaxation at different base temperature provides about 1000 data 
points over a temperature interval of up to 50\,\% above the base temperature. 
Magnetization was measured with a commercial Quantum Design Vibrating-Sample 
Superconducting Quantum Interference Device (VSM-SQUID).

\begin{figure}[t]
\begin{center}
\includegraphics[width=85mm]{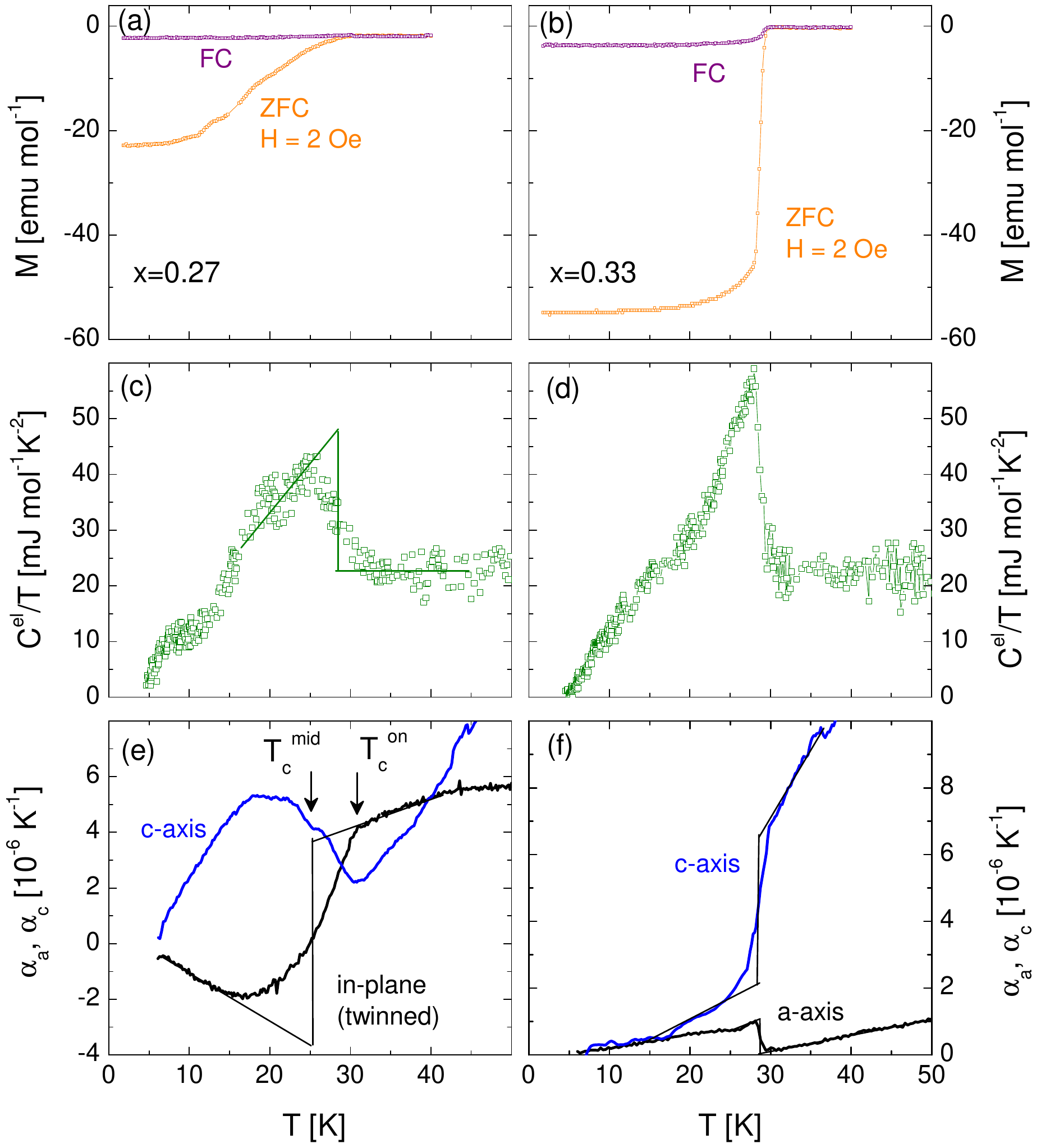}
\end{center}
\caption{The superconducting transition of samples with $x=0.27$ (underdoped, 
left column) and $x=0.33$ (overdoped, right column) seen by different probes. 
Magnetization (panels (a) and (b)), specific heat (panels (c)-(d)) and thermal 
expansion (panels (e)-(f)) all show a broad superconducting transition with a 
sharp onset for the underdoped sample and more standard, sharp anomalies for the 
overdoped sample.}
\label{fig:transition}
\end{figure}

\section{Results}\label{sec:results}

Fig. \ref{fig:alphaall} shows the measured uniaxial thermal-expansion 
coefficients $\alpha_i=(1/L_i) d\!L_i/d\!T$ where $L$ is the sample length and the 
subscript $i$ stands for the direction. Clear anomalies are observed at the 
magneto-structural transition (at $T_{sm}$) and at the onset of 
superconductivity at $T_c^{\ on}$ as indicated in the Figure. Panel (b) shows the 
detwinned measurements of underdoped samples, which become orthorhombic upon cooling.
It is clear that the process of detwinning strongly enhances all anomalies when compared 
to the twinned measurements (panel (a)). The magneto-structural transition 
manifests itself as a broadened peak. Panel (e) shows how this peak can actually be 
decomposed into two separate peaks, revealing a small splitting of the structural transition at $T_s$ and 
the magnetic transition at $T_{SDW}$. 
We find $T_s-T_{SDW}\approx2.5\,$K for $x=0.18$ and $\sim7\,$K for $x=0.25$ by careful comparison of the different in-plane measurements. 
For $x=0.27$ no clear splitting of the transition can be observed.
Samples with $x\geq0.25$ show bulk superconductivity. The 
onset of superconductivity on the underdoped side is signaled by a rather sharp 
kink (at $T_c^{\ on}$) followed by a broad peak, which has the opposite sign of the anomaly at the 
structural transition. We will use an area-conserving construction (see panel (e) 
of Fig. \ref{fig:transition}) to define a $T_c^{\ mid}$. On the overdoped side 
($x=0.33$), the anomaly at $T_c$ has the usual, step-like, shape expected at a second-order phase transiton, and 
$T_c^{\ on}$ and $T_c^{\ mid}$ nearly coincide. Note that the sign of the thermal-expansion 
anomaly allows us to assign a sample uniquely to lying either on the underdoped 
or on the overdoped side. The anomalies of the thermal-expansion coefficient along 
the $c$-axis (panel (c)) have similar shape but opposite sign as compared to the 
in-plane measurements in all cases. It is important to note that the anomalies of the 
$\alpha_i$'s have the same sign and exhibit a very similar shape as reported for the 
Ba(Fe$_{1-x}$Co$_x$)$_2$As$_2$ (Co-Ba122) system\cite{Budko2009, Hardy2009, 
Meingast2011}. Since $\alpha_i\propto -d\!S/d\!p_i$ (uniaxial pressure derivatives of the 
entropy), this shows that the entropy of the two systems responds similarly to 
uniaxial pressure.

The strikingly different shape of the superconducting transition of 
underdoped and overdoped samples is also seen in other measurements. 
Fig. \ref{fig:transition} shows the superconducting transition of the underdoped 
($x=0.27$) and overdoped ($x=0.33$) P-Ba122 samples in the dc-magnetization, the 
electronic specific heat (derived by subtracting the lattice 
heat capacity of an undoped Ba122 sample), and the thermal expansion. For underdoped 
BaFe$_2$(As$_{0.73}$P$_{0.27}$)$_2$ there is a sharp onset at $T_c^{\ on}$ in all three data sets, however the 
main anomaly appears very broad and rounded. The approximate transition width is $10-15\,$K. On the other hand, overdoped 
BaFe$_2$(As$_{0.67}$P$_{0.33}$)$_2$ exhibits sharp, step-like anomalies  of 
$\alpha$, $C_p$ and $M$ at $T_c$. A possible origin of the broadening in the underdoped samples 
will be presented in Section \ref{sec:phasediagram}.

\section{Phase diagram}\label{sec:phasediagram}

\begin{figure}[t]
\begin{center}
\includegraphics[width=85mm]{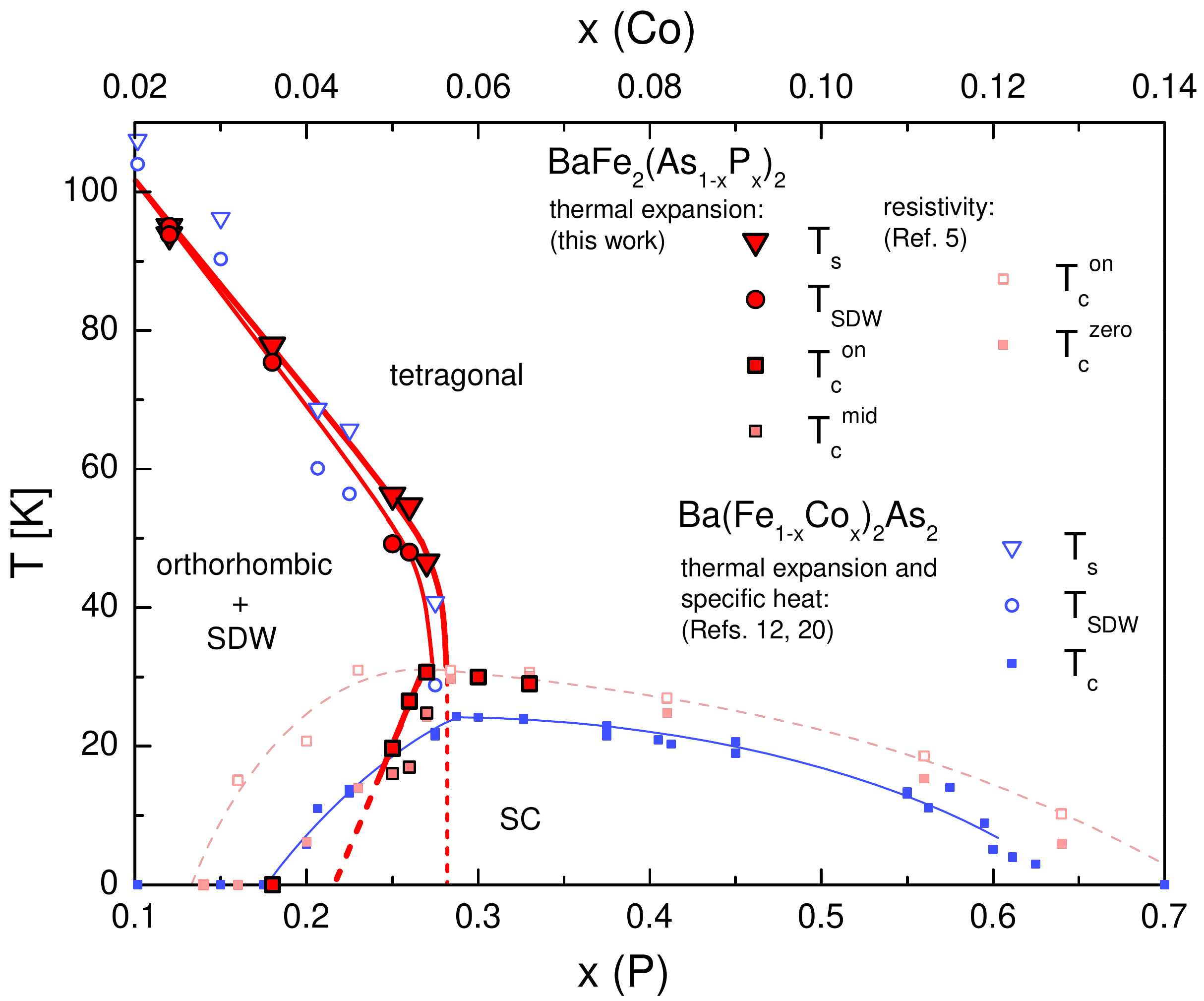}
\end{center}
\caption{Thermodynamic phase diagrams of P-Ba122 (lower axis) compiled from the present 
thermal-expansion data and of Co-Ba122 (upper axis) scaled so that the maximum $T_c$'s coincide. Data for Co-Ba122 was 
obtained from specific-heat\cite{Hardy2010II} and thermal-expansion\cite{Meingast2011} measurements. 
Also added are previously published resistivity data\cite{Kasahara2010} on the P-Ba122 system. 
Clearly, the bulk $T_c$ rises more steeply for P-Ba122 than for Co-Ba122 on the underdoped side. }
\label{fig:phasediag}
\end{figure}

Fig. \ref{fig:phasediag} shows the phase diagram compiled from the present thermodynamic data
together with previous work on resistivity\cite{Kasahara2010} of the same system and thermodynamic 
data for Co-Ba122 (Refs. \onlinecite{Hardy2010II, Meingast2011}), 
which have been scaled so that the optimal doping concentrations of both systems coincide. 
In underdoped P-Ba122, the bulk $T_c$ obtained from thermal expansion is lower 
than $T_c$ inferred from resistivity. Furthermore, the thermodynamic measurements 
reveal a very steep slope of $T_c$. This may offer a simple 
explanation for the broad superconducting transitions of the
underdoped samples of P-Ba122 (Fig. \ref{fig:transition}). For a given concentration gradient 
in the sample, the width of the transition will be directly related to this slope. For $x=0.25-0.27$ 
the width of $T_c$ amounts to $10-15$\,K while it is only $\sim 1\,$K on the 
overdoped side. This difference can thus be explained by the roughly $13$ times greater slope of $T_c$ 
on the underdoped side. However, the sharp kinks at $T_c^{\ on}$ remain 
unexplained. We note that quite broad superconducting transitions on the 
underdoped side have also been reported previously from resistivity\cite{Kasahara2010} 
and susceptibility\cite{Iye2012} measurements. 

We have already noted that the response of the entropy to uniaxial pressure is 
very similar for P-Ba122 and Co-Ba122. Overall their phase diagrams are also very similar, 
however, they also show some important differences. First, on the underdoped side, the bulk
$T_c$ rises more steeply in P-Ba122 than in the Co-doped system. This may be a signature of a stronger 
competition between the SDW and the superconducting phase. A steep rise of the 
transition line signals that two phases differ little in entropy so that only a 
significant increase of temperature triggers the transition. This means that here, superconductivity 
is not much ``weaker'' than magnetism, and a strong effect of the two types of order on each other
is expected. 

A second difference of the phase diagrams of P-Ba122 and Co-Ba122 is that the splitting 
of the magneto-structural transition is about two times smaller in the P substituted system.
Interestingly, the two transitions remain coincident for isovalent Ru 
substitution\cite{Thaler2010}. This suggests that, surprisingly, isovalently substituted P-Ba122 is closer to Co-Ba122 than to Ru-Ba122. 
This may be connected to the proposition that also Co substitution does not induce charge doping\cite{Wadati2010, Merz2012}.
The reduced magnitude of the splitting may be explained by less scattering in the P-substituted samples\cite{vanderBeek2010}. 
A reduction of the splitting with decreasing disorder has been reported previously 
in the 1111 systems\cite{Jesche2010}. Finally, on the overdoped side, $T_c$ is uniformly higher for P substitution 
than for Co substitution, possibly also due to less disorder.

\section{Interplay of orthorhombicity, SDW and superconductivity}\label{sec:interplay}

\begin{figure}[t]
\begin{center}
\includegraphics[width=70mm]{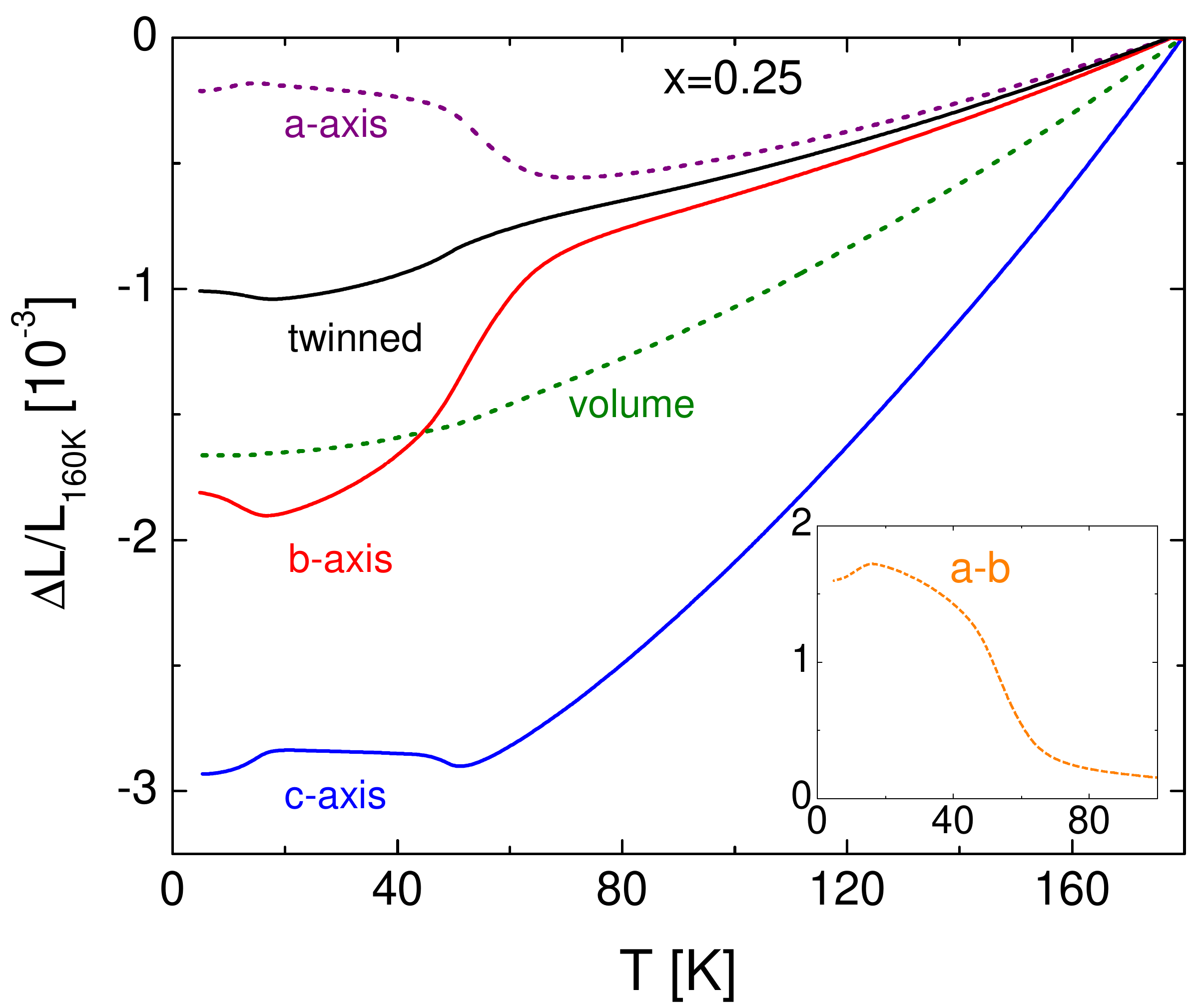}
\end{center}
\caption{Relative change of sample length of the 
BaFe$_2$(As$_{0.75}$P$_{0.25}$)$_2$ sample, measured in the twinned (in-plane average, black line) and detwinned (b-axis, red line) configuration and along the $c$-axis (blue line). The $a$-axis length (broken purple line) was estimated from the twinned and detwinned in-plane measurements and the volume change (green broken line) from the twinned in-plane and the $c$-axis measurements. The inset shows the difference of $a$-axis and $b$-axis. For this sample $T_{sm}\approx55\,$K and $T_c^{\ mid}=16\,$K. }
\label{fig:intalphaabc}
\end{figure}

Recent studies on Co-Ba122 showed that both magnetic and structural order parameters 
are reduced in the superconducting state\cite{Pratt2009,Christianson2009,Nandi2010}.
This has been interpreted as evidence for microscopic coexistence and competition
 between the orthorhombic SDW phase and superconductivity 
in the underdoped region of the phase diagram\cite{Pratt2009,Fernandes2010III}.  
In some of the 1111 compounds, on the other hand, no such coexistence region has been observed\cite{Luetkens2009}, and it is still debated 
whether magnetism and superconductivity coexist microscopically in K-doped Ba122\cite{Park2009,Inosov2009,Wiesenmayer2011}.  
It is thus of interest to see whether such a coexistence also occurs in P-Ba122.  
A very recent NMR study\cite{Iye2012} shows evidence for a reduced magnetic order parameter in the superconducting 
state also for P-Ba122 and therefore claims microscopic coexistence of magnetism and superconductivity. In the 
following we will use our thermal-expansion data to study this interplay between structural/magnetic order and 
superconductivity via the response of the samples' lengths and the orthorhombic order parameter to superconductivity. 

As an example, Fig. \ref{fig:intalphaabc} shows the relative length changes of
the $x=0.25$ sample. We note that the $a$-axis length was not directly measured.  
However, it can be derived from the difference between in-plane twinned and detwinned data. 
The difference of the relative length changes of $a$- and $b$-axes, 
i.e., the orthorhombic order parameter $\delta(T)$ inferred from our measurements, is shown in the inset of Fig. \ref{fig:intalphaabc}.
We see that  $a$- and $b$-axis lengths start to differ clearly at the structural transition, 
and this difference is directly proportional to the structural order parameter in this system. 
There exists a high-temperature tail to the transition, which arises from the small applied 
in-plane pressure of the order of $5-10$\,MPa\cite{Blomberg2011,Dhital2011}. Clearly, $a$- and $b$-axes approach each other 
again below $T_c$ and the orthorhombic order parameter decreases, 
suggesting a similar coupling of structural and superconducting order parameters 
as observed in Co-Ba122 (Ref. \onlinecite{Nandi2010}). 
Interestingly, the $c$-axis length increases below $T_s$
and then decreases below $T_c$, and thus exhibits a very similar behavior as the in-plane axes.  
Although it is not clear how, or if at all, the $c$-axis response can be linked to the structural/magnetic 
order parameters, our results directly show that the SDW state favors a longer $c$-axis. 
Equivalently, compressing the $c$-axis by uniaxial pressure will destabilize magnetism. This is in agreement 
with DFT calculations under uniaxial pressure\cite{Tomic2011}. 
The decrease of the $c$-axis length upon entering the superconducting 
state is thus consistent with a suppression of magnetism by superconductivity.  
Finally, we note that the effect of these transitions on the volume is very small (see Fig. \ref{fig:intalphaabc}) 
due to an almost complete cancellation of the anomalies along the different directions.

\begin{figure}[t]
\begin{center}
\includegraphics[width=85mm]{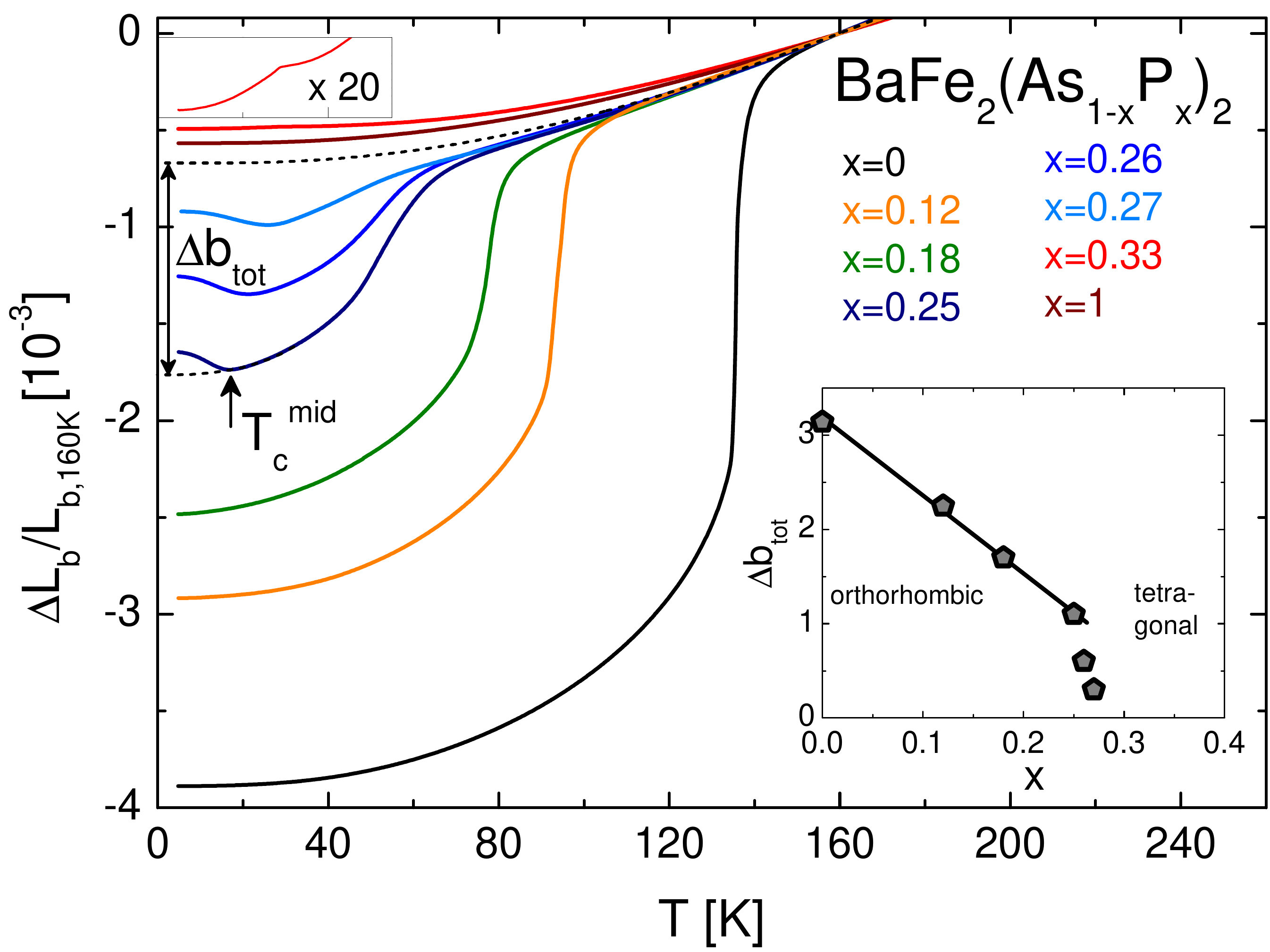}
\end{center}
\caption{Relative change of the $b$-axis length (in-plane length for the overdoped $x=0.33$ sample) for various doping levels. Its 
shortening signals the structural transition and its increase below 
$T_c$ for the superconducting samples demonstrates the competition between the 
orthorhombic phase and superconductivity. The upper inset shows the data for $x=0.33$ on an expanded vertical scale. The lower inset shows the doping evolution 
of $\Delta\! b_{tot}$ as defined in the main panel.}
\label{fig:deltab}
\end{figure}

\begin{figure}[t]
\begin{center}
\includegraphics[width=90mm]{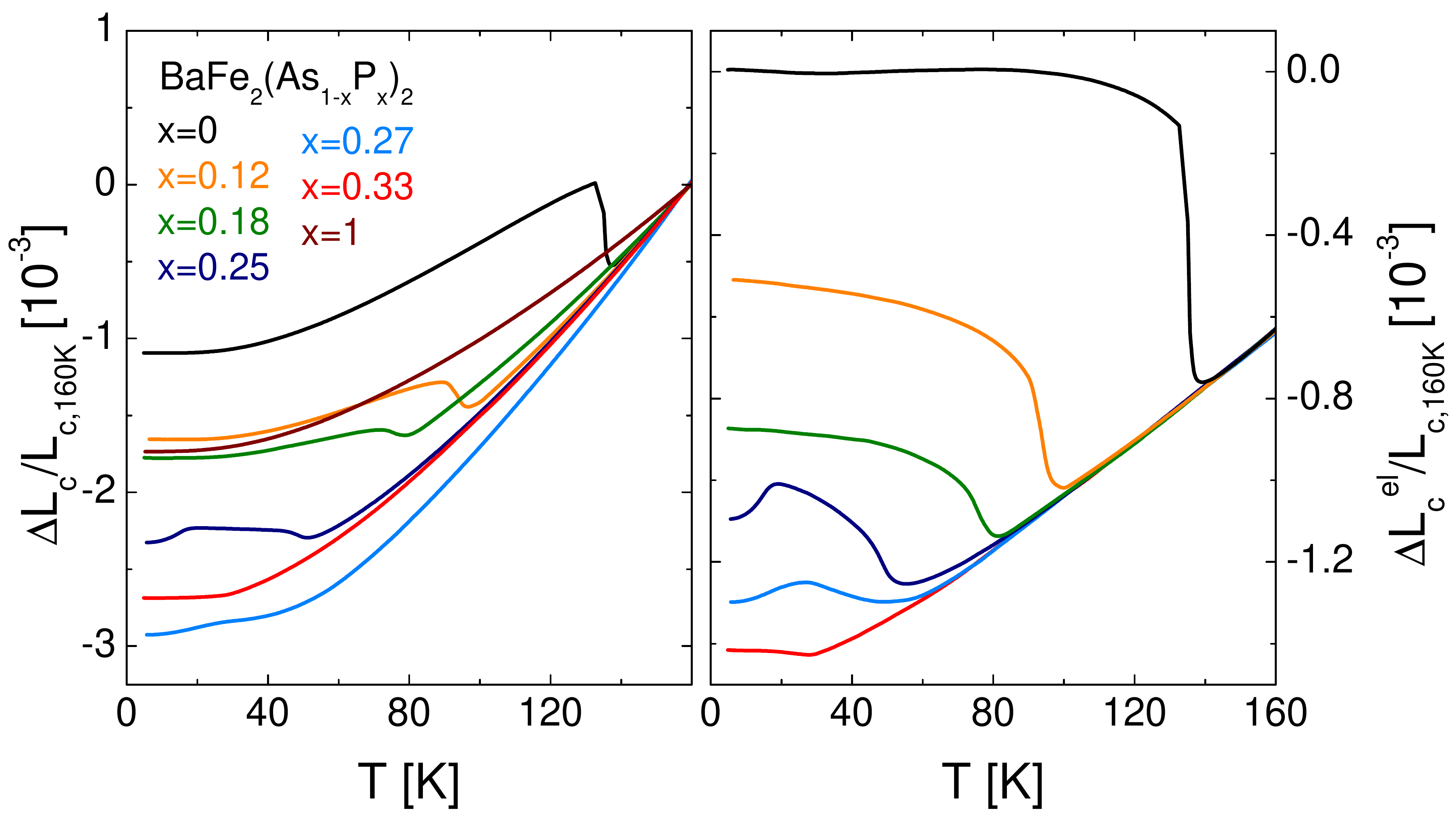}
\end{center}
\caption{(a) Relative change of the $c$-axis length for various doping levels. 
Panel (b) shows the data after subtraction of a background (the data for $x=1$ times an individual factor close to 1, see text). }
\label{fig:DeltaLc}
\end{figure}

In Figures \ref{fig:deltab} and \ref{fig:DeltaLc} we show the 
length changes as a function of temperature for different doping levels 
for the $b$- and $c$-axes, respectively. These axes can be measured accurately. 
We do not show the twinned and inferred $a$-axis data or $\delta(T)$,  
since these are not as reliable due to possible partial 
detwinning during the twinned in-plane measurements. 
We note that the temperature dependence of the orthorhombic order parameter $\delta=(a-b)/(a+b)$ is 
however reflected in the (negative) $b$-axis length since $a$- and $b$- axis are found 
to evolve nearly symmetrically. In Fig. \ref{fig:deltab} we define a parameter $\Delta b_{tot}$, 
which quantifies the total change of the $b$-axis length due to the magneto-structural 
 transition. As shown in the inset of this figure, $\Delta b_{tot}$ 
 decreases linearly with P content confirming reliable detwinning. An exception are the samples with $x=0.26$ and $0.27$. 
 Even though the transition temperatures are similar, $\Delta b_{tot}$ is much smaller 
 for these samples and does not follow $T_s$ any more. Possibly, only a part of the 
sample undergoes the magneto-structural transition. Similarly, NMR found evidence that a 
fraction of a collection of samples with $x=0.25$ does not undergo the SDW 
transition\cite{Iye2012}. Hence we exclude the samples with $x=0.26$  and $0.27$ in the 
analysis of the electronic part of the thermal expansion (Section \ref{sec:pressureeffects}).
As already seen for the $x=0.25$ sample, the $b$-axis length increases and the 
orthorhombicity is significantly reduced when samples become superconducting.
On the other hand, the in-plane length of the overdoped $x=0.33$ sample decreases 
below $T_c$. However, the effect is tiny in comparison to the underdoped samples 
(see inset in the upper left corner of Fig. \ref{fig:deltab}).

Interestingly, the $c$-axis length displays a similar behavior as the in-plane axes.
Fig. \ref{fig:DeltaLc} (a) shows length changes of the $c$-axis
in analogy to Fig. \ref{fig:deltab}. The iron-based materials are more 
compressible along the $c$-axis than in-plane, which is reflected in a larger 
lattice contribution to the thermal expansion along the $c$-axis\cite{Hardy2009,Meingast2011}. 
To obtain the ``electronic $c$-axis length changes'' $\Delta\!L_c^{el}$ a suitable lattice background has to be subtracted. 
For this lattice background we used the $x=1$ data, because pure BaFe$_2$P$_2$ does not undergo
any phase transitions and, additionally, the electronic contribution to its thermal expansion is negligibly small
compared to the effects discussed here. We assume that this background is independent of P doping and, furthermore, 
that the high-temperature value of the total $\alpha_c$ is doping independent. 
Variations of this high-temperature value are attributed to uncertainties arising from the tiny sample lengths along $c$ ($\sim70-150\,\mu$m).
These deviations can be modeled with a correction factor between $0.9$ and $1.3$ to the lattice contribution 
for the different samples. This correction does not affect the shape of the anomalies.  
The thus obtained $\Delta\!L_c^{el}$ are shown in panel (b) of Figure \ref{fig:DeltaLc} as a function of temperature. 
Intriguingly, these $\Delta\!L_c^{el}$ resemble closely the $T$ dependence 
of the orthorhombic order parameter for all doping levels. This opens the interesting possibility 
to study the orthorhombic SDW state via its $c$-axis length. As already detailed in the 
analysis of the $x=0.25$ sample, it is clearly seen that the magnetic state 
favors a longer $c$-axis. Its destabilization by superconductivity is reflected 
in a reduced $c$-axis length. Interestingly, along the $c$-axis, there is a marked 
response also for the overdoped $x=0.33$ sample. The superconducting state in the absence of static magnetic order 
apparently favors a longer $c$-axis in a similar manner as does the SDW state of 
underdoped P-Ba122. The same effect, though much smaller in magnitude, is also seen for the in-plane dimension: 
The $a$-axis length of overdoped P-Ba122 is (slightly) reduced 
in response to superconductivity, as is the average in-plane length of the underdoped material 
upon entering the SDW state. 
The similar response of the lattice parameters may hint at a kinship of SDW and superconductivity 
as is suggested by nearly identical Grüneisen-parameters of the two phases in Co-Ba122\cite{Meingast2011}.  

\begin{figure*}[t]
\begin{center}
\includegraphics[width=180mm]{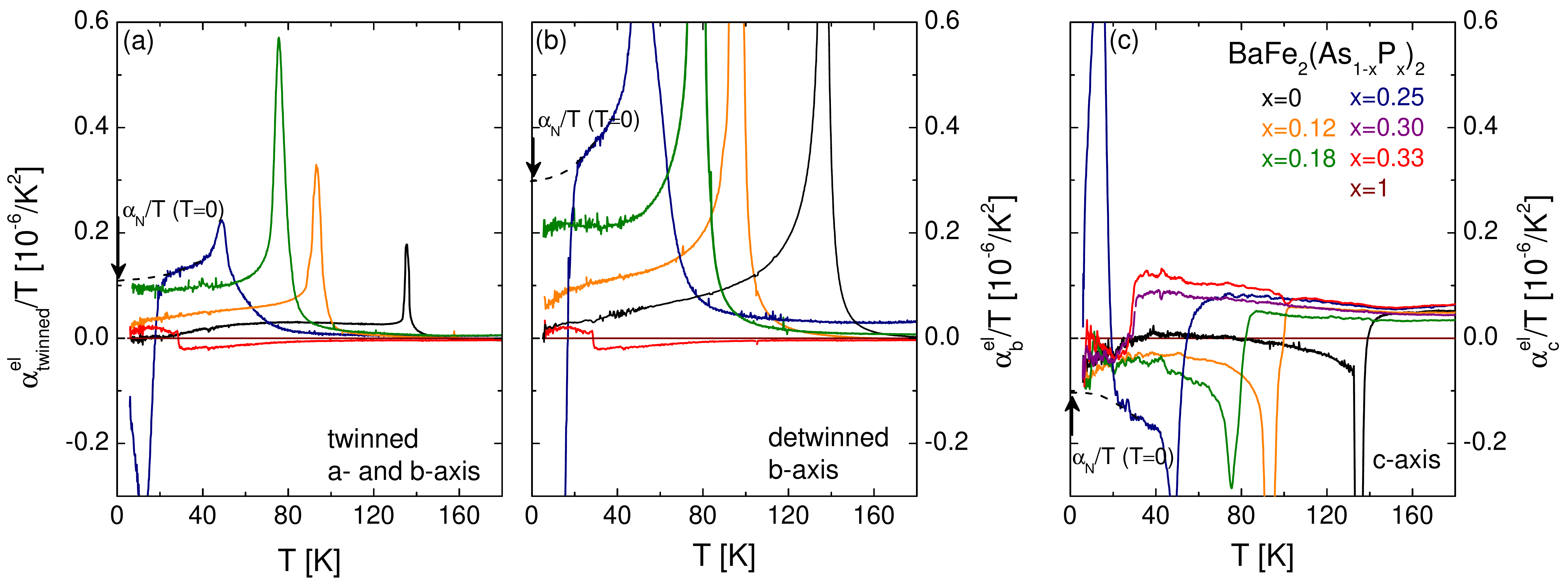}
\end{center}
\caption{The electronic contribution to the uniaxial thermal-expansion 
coefficients divided by $T$, $\alpha_{el}/T$, obtained from subtracting the data of the $x=1$ sample as a phonon 
background (a) in the twinned in-plane measurements, (b) along the orthorhombic 
$b$-axis (detwinned in-plane measurements) and (c) along the $c$-axis. Arrows indicate how values of the normal state
$\alpha_{i}^{el}/T (T=0)$, which are reported in Fig. \ref{fig:dgammadp}, are extracted.}
\label{fig:alphaelectronic}
\end{figure*}

Finally, we address the question of microscopic coexistence of magnetism and superconductivity. 
Dilatometry is a macroscopic probe and the $T$ dependence of a sample length cannot \textit{a priori} be equated 
with the $T$ dependence of the lattice parameters if one has a phase-separated sample. Still, the close resemblance between 
our results and the evolution of lattice parameters of Co-Ba122 observed by x-ray diffraction\cite{Nandi2010} 
supports that such an identification may indeed be done, suggesting that our data really show the effect 
of superconductivity on the lattice constants. This effect is most naturally explained by homogeneous 
coexistence and competition of magnetism and superconductivity also in P-Ba122. In confirmation, 
NMR measurements\cite{Iye2012} show an influence of superconductivity on the local magnetic order 
parameter and hence also conclude that SDW and superconductivity coexist microscopically but compete with each other.  

Summarizing this section, the temperature and doping dependence of the lattice parameters of 
underdoped P-Ba122 reflect how superconductivity competes with the orthorhombic magnetic state and finally suppresses it. 
In contrast, superconductivity on the overdoped side is stabilized by lattice changes similar to
those that occur upon entering the SDW phase, which hints at a close similarity of the two ordering phenomena.  

\section{Uniaxial pressure effects}\label{sec:pressureeffects}

For second-order phase transitions, uniaxial pressure derivatives of the
transition temperature, $d\!T_c/d\!p_i$, can be deduced from the jumps
of the
thermal-expansion coefficient $\Delta\!\alpha_i$ and the specific heat
$\Delta\!C_p$. For the overdoped, tetragonal, sample ($x=0.33$, $T_c=29\,$K), clear second-order jumps
in the uniaxial thermal-expansion coefficients
$\Delta\!\alpha_a=0.9(2)\times10^{-6}$K$^{-1}$,
$\Delta\!\alpha_c=-4.0(5)\times10^{-6}$K$^{-1}$
 and the specific heat $\Delta\!C_p/T_c=38(4)\,$mJ mol$^{-1}$ K$^{-2}$ are
observed (see Fig. \ref{fig:transition}). The magnitude of the specific-heat jump
is in good agreement with data of a recent report by Chaparro et al. \cite{Chaparro2011}.
$d\!T_c/d\!p_i$ can be calculated via the Ehrenfest relation
\begin{equation}
\frac{d\!T_c}{d\!p_i}=V_m\frac{\Delta\!\alpha_i}{\Delta C_p/T_c}.
\end{equation}
where $V_m=59.6$\,cm$^3$/mol (Ref. \onlinecite{Kasahara2010}) is the molar volume. The data
yield $d\!T_c/d\!p_a=1.4(5)$\,K/GPa and $d\!T_c/d\!p_c=-6.3(1.2)$\,K/GPa. The
hydrostatic pressure derivative $d\!T_c/d\!p=2d\!T_c/d\!p_a+d\!T_c/d\!p_c=-3.5(2.2)$\,K/GPa 
is in reasonable agreement with high-pressure experiments
\cite{Goh2010} which report an initial slope of
$d\!T_c/d\!p=-1.8$\,K/GPa for a
sample with $T_c$=30.5\,K. The same kind of anisotropy and a similar
magnitude of the uniaxial pressure derivatives have been observed for a slightly
overdoped Co-Ba122 crystal \cite{Hardy2009}. 
The pressure derivatives on
the underdoped side are more difficult to extract due to the large
widths of the transitions.  However, it is clear that the $d\!T_c/d\!p_i$'s are of
opposite sign and by far larger in magnitude than those of the overdoped sample.
Using the construction for the jumps shown in Fig. 2(e), we find for the
$x=0.27$ sample $d\!T_c/d\!p_{ab}=-18(8)$\,K/GPa,
$d\!T_c/d\!p_{b}=-44(17)$\,K/GPa  and $d\!T_c/d\!p_c=20(8)$\,K/GPa. Here, $p_{ab}$ refers to an average of pressure 
along the $a$- and $b$-axis, as is relevant to the twinned in-plane measurements.  

Since $\Delta\!C_p$ is always positive, the sign of the anomalies in the 
$\alpha_i$ provide information on the anisotropy of the $d\!T_c/d\!p_i$ even 
in the absence of specific-heat data. Strikingly, the uniaxial pressure 
derivatives of $T_{s}$ and $T_{SDW}$ have the opposite sign to the derivatives of $T_c$ for 
all directions\cite{Boehmer2011}. This is again consistent with a competition between 
the orthorhombic SDW phase and superconductivity, since pressure will
favor either the SDW or superconducting phase at the expense of
the other. Interestingly, all signs of 
the derivatives can be accounted for by identifying uniaxial pressure with a 
shift in the P content: stress along the $c$-axis (and $a$-axis on the 
underdoped side) corresponds to an increased P content while stress along the 
$b$-axis (averaged in-plane axis) corresponds to a lower P content. This means 
that the phase diagram may be tracked (forwards or backwards) by the application of uniaxial 
pressure just as it can be tracked by hydrostatic pressure\cite{Klintberg2010}. 
By comparing the values of $d\!T_c/d\!p_{i}$ obtained above with the respective 
slopes in the phase diagram $d\!T_c/d\!x=6.7\,$K/at\%(P) (underdoped) and $d\!T_c/d\!x=-0.5\,$K/at\%(P) (overdoped) 
one can estimate that increasing the P content by 1\,at\% corresponds to a uniaxial pressure 
of $-0.35$\,GPa ($-0.15$\,GPa) in the in-plane average (along the $b$-axis). 
The lower accuracy of the $c$-axis data unfortunately does not allow such a quantitative analysis.  

\begin{figure}[t]
\begin{center}
\includegraphics[width=75mm]{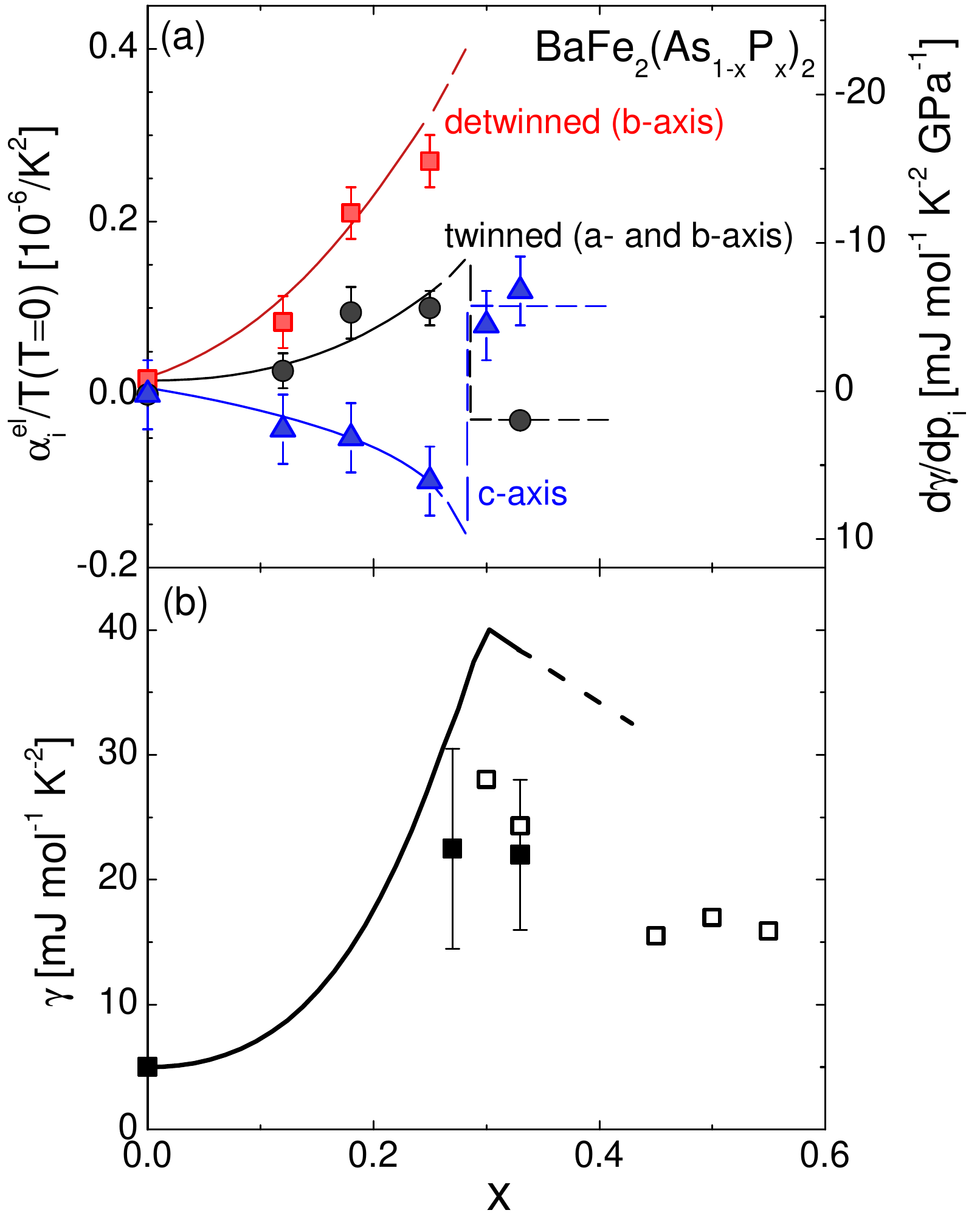}
\end{center}
\caption{(a) Values of the electronic thermal expansivity $\alpha_{el,i}/T$ extrapolated to zero 
temperature. The right-hand scale shows the corresponding values of 
$d\!\gamma/d\!p_i$. Lines are a guide to the eye. (b) The Sommerfeld coefficient $\gamma$ obtained by integrating the smooth lines from (a) (black line), see text for details. Full squares indicate directly measured values of $\gamma$ and open squares indicate values estimated from the measured specific-heat jump\cite{Chaparro2011}, see also note \onlinecite{gammavalues}.}
\label{fig:dgammadp}
\end{figure}

Thermal expansion data can also be used to obtain information about
the (uniaxial) pressure dependence of the electronic density of states.
For a Fermi liquid, the electronic thermal expansion divided by $T$,
$\alpha_{i}^{el}/T$, is expected to be a constant, $\alpha_{el,i}/T=-(1/V_m)d\!\gamma/d\!p_i$, 
with $\gamma$ the Sommerfeld coefficient. (This follows
directly from $\alpha_i=-(1/V_m)d\!S/d\!p_i$ and the electronic specific
heat $C=\gamma T$.)  Fig. \ref{fig:alphaelectronic} shows the
electronic/magnetic thermal-expansion coefficients $\alpha_{i}^{el}/T\propto -d\!\gamma/d\!p_i$, which were
obtained from the original data by subtracting the data for $x=1$ as a phonon background 
using no correction factor (see section \ref{sec:interplay}).  With increasing doping, a sizeable
contribution from electronic/magnetic degrees of freedom evolves below the SDW transition
for all three crystal directions. Especially for the $x=0.18$ data,
this contribution is nearly independent of temperature up to about $60\,$K,
consistent with Fermi-liquid-type behavior. This is reminiscent of
what is observed in weak itinerant ferromagnetic systems, such as MnSi (Ref. \onlinecite{Matsunaga1982}), 
and points to an intricate mixing of highly pressure-dependent
electronic and magnetic degrees of freedom. For $x=0.25$, $T_{SDW}$ and $T_c$
are too close to each other so that this constant term in $\alpha_{el}/T$ is not seen.

Panel (a) of Fig. \ref{fig:dgammadp} shows $\alpha_{i}^{el}/T(T=0)$, obtained by
extrapolating the normal state $\alpha_{i}^{el}/T$ data to zero, 
and corresponding $d\!\gamma/d\!p_i$ values versus P substitution. There is a sharp increase in
the magnitude of $d\!\gamma/d\!p_i$ upon approaching
optimal doping and a sign change when passing to the overdoped side. The discontinuity of the derivatives $d\!\gamma/d\!p_i$ at optimal doping 
is also suggested from an analogy with the Co-doped system\cite{Hardy2010II, Meingast2011}. We can obtain $\gamma$ as a
function of $x$ from integration of these data if, as before, we equate doping with
pressure. We recall that we can link average in-plane ($c$-axis) pressure to a decreased (increased) P content. As is evident from the 
sign change then, $\gamma(x)$ has a maximum at optimal doping. More quantitatively, the proportionality factor from above   
($d\!p_{ab}/d\!x =-0.35\,$GPa/at\%(P) and $d\!p_{b}/d\!x =-0.15\,$GPa/at\%(P)) yields values 
for $d\!\gamma/d\!x$ as a function of doping; for example 
$d\!\gamma/d\!x=2.4\,$mJ mol$^{-1}$K$^{-2}$/at\%(P) at $x=0.25$. Integration of the smooth 
lines in Fig. \ref{fig:dgammadp}(a) rescaled by this factor results in $\gamma(x)$ shown in Fig. \ref{fig:dgammadp}(b). 
$\gamma$ increases with doping in the underdoped region, has a sharp maximum at optimal doping and
then decreases with further doping. Also shown are measured values\cite{gammavalues} of $\gamma$, which agree reasonably well
considering the assumptions and uncertainties of our approach and the uncertainties in the values of $\gamma$. The
overall behavior of $\gamma(x)$ is very similar to what has been observed
in Co-doped Ba122, where the initial increase in $\gamma$ was argued to
result from the gradual suppression of the SDW order with Co doping\cite{Hardy2010II,Meingast2011}.
Clearly, similar physics is at work in the P-doped system. On the other hand, a sharp increase of $\gamma$ 
would also be expected from the presence of a quantum critical point close to optimal doping.

\section{Conclusions}\label{sec:conclusions}
We have constructed the thermodynamic phase diagram of P-Ba122 in a 
significant range of P substitution $0\leq x \leq0.33$. Compared to the Co-substituted system, the superconducting $T_c$ rises 
more steeply and there is only a small overlap region where
superconductivity occurs within the orthorhombic SDW phase. P-Ba122 seems to be rather close to a first-order 
transition between magnetism and superconductivity, as is realized in some of the 1111 systems. 
The question arises why P-Ba122 and Co-Ba122 differ in this aspect of interplay between SDW and superconductivity.
The competition of the two types of order is also reflected in the lattice constants. 
The destabilization of the SDW by 
superconductivity is reflected both in the in-plane lengths, as a reduction of orthorhombicity, 
and in the $c$-axis length. However, superconductivity on its own favors a change 
of lattice parameters similar to the SDW state, suggesting an intimate relationship of the two ordering phenomena. 
Uniaxial pressure derivatives of the transition temperatures are large and anisotropic. 
Considering the effects on the phase diagram, we put forward a linear relation between uniaxial 
pressure and P content. The Sommerfeld coefficient is increasingly pressure 
(and therefore doping) dependent on approaching optimal doping. 
By evaluating its derivatives, we deduce that the density of states 
has a maximum at optimal doping.

\begin{acknowledgements}
We cordially thank J. Schmalian, T. Iye and I. Eremin for discussions. This work was 
supported by the Deutsche Forschungsgemeinschaft through SPP 1458, 
by the Research Grants Council of Hong Kong, Grants 603010 and SEG\_HKUST03 and by
Grant-in-Aid for GCOE program ``The Next Generation of Physics, Spun from Universality
and Emergence'' from MEXT, Japan. 
\end{acknowledgements}

\end{document}